\documentclass[12pt]{article}
\usepackage{jeosman}
\usepackage{epsfig}
\usepackage{amsmath}
\usepackage{amssymb}

\begin{document}

\title{Slow-light enhanced absorption for bio-chemical sensing applications: potential of low-contrast lossy materials}

\author{Jesper Pedersen, Sanshui Xiao, and Niels Asger Mortensen}

\address{MIC -- Department of Micro and Nanotechnology, NanoDTU,\\Technical
University of Denmark, DK-2800 Kongens Lyngby, Denmark}

\email{asger@mailaps.org}


\begin{abstract}
Slow-light enhanced absorption in liquid-infiltrated photonic
crystals has recently been proposed as a route to compensate for
the reduced optical path in typical lab-on-a-chip systems for
bio-chemical sensing applications. A simple perturbative
expression has been applied to ideal structures composed of
lossless dielectrics. In this work we study the enhancement in
structures composed of lossy dielectrics such as a polymer. For
this particular sensing application we find that the material loss
has an unexpected limited drawback and surprisingly, it may even
add to increase the bandwidth for low-index contrast systems such
as polymer devices.
\end{abstract}

\maketitle

\section{Introduction}

The strong emphasis on miniaturization of chemical analysis
systems~\cite{Janasek:2006} has naturally fuelled a large effort
in integrating optics and microfluidics in lab-on-a-chip
microsystems~\cite{Verpoorte:2003,Mogensen:2004} and more recently
the emerging field of
optofluidics~\cite{Psaltis:2006,Monat:2007a,Erickson:2008} has
increased this interest further.

The miniaturization of chemical analysis systems has created a new
paradigm with a broad variety of phenomena, properties, and
applications benefitting from the down-scaling in
size~\cite{Janasek:2006}. However, the case of optical sensing and
detection remains an exception because the light-matter
interactions suffer from the reduced optical path length $L$ in
lab-on-a-chip systems compared to their macroscopic counterparts.
For Beer--Lambert absorption measurements, a typical size
reduction by two orders of magnitude will penalize the optical
sensitivity in an inversely proportional manner as has been
demonstrated by Mogensen \emph{et al.}~\cite{Mogensen:2003}.

\begin{figure}[t!]
\begin{center}
\epsfig{file=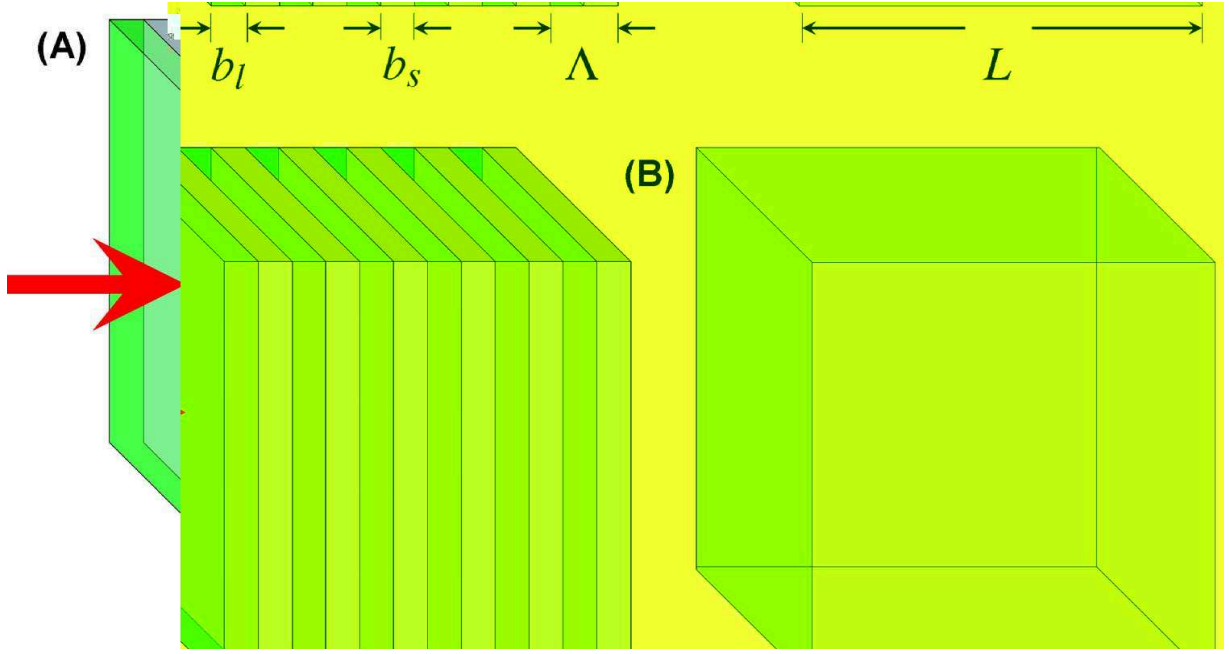, width=\columnwidth,clip}
\end{center}
\caption{(A) Liquid-infiltrated Bragg stack composed of
alternating solid and liquid layers of thickness $b_s$ and $b_l$,
respectively. (B) Schematic illustration of a corresponding
classical Beer--Lambert absorbance cell where the incident light
passes through a spatially homogeneous medium, i.e. the liquid
sample to be analyzed.} \label{fig1}
\end{figure}

In a simple picture, the above problem stems from the reduced
light-matter interaction time $\tau$. For a homogeneous system of
length $L$, such as a classical absorbance cell, we have the
following Wigner--Smith delay time
\begin{equation}\label{eq:tau}
\tau\sim L/c,
\end{equation}
with $c$ being the speed of light. Obviously, for small systems
the photons spend too little time inside the sample and thus do
not inherit strong fingerprints from the chemical species they
were supposed to interact with. As a result, only very high
concentrations can be quantified compared to the levels that can
be detected with a macroscopic optical path length. On the other
hand, lab-on-a-chip systems are often envisioned for applications
dealing with quite minute sample volumes with very low
concentrations of bio-molecules. The use of slow-light phenomena
in liquid-infiltrated Bragg stacks and photonic crystals, see
Fig.~\ref{fig1}, has recently been proposed as a way to deal with
this conflict of length
scales~\cite{Mortensen:2007a,Pedersen:2007a}.

In the context of Beer--Lambert absorption, the modulation of the
transmitted intensity is typically expressed as
\begin{equation}
I=I_0 \exp\left(-\alpha L\right)=I_0 \exp\left(-\gamma \alpha_l
L\right)
\end{equation}
where $\alpha_l$ is the absorption coefficient for the absorbing
liquid and $I_0$ is the transmitted intensity with a vanishing
concentration of the absorbing bio-chemical species in the liquid,
while $I$ is then the transmitted intensity at a finite
concentration which is to be quantified by the absorbance
measurement. To lowest order in the absorption, the dimensionless
enhancement factor $\gamma\equiv \alpha/\alpha_l$ can be expressed
as~\cite{Mortensen:2007a}
\begin{equation}
\label{eq:gamma} \gamma =f\times\frac{c/n_l}{v_g},
\end{equation}
where $0<f<1$ is a dimensionless number quantifying the relative
optical overlap with the liquid~\cite{Mortensen:2008a}. The
fraction on the right-hand side expresses the ratio of the group
velocity $c/n_l$ in the bare liquid to the group velocity $v_g$ in
the liquid-infiltrated structure, thus clearly illustrating the
enhancement by slow-light propagation ($v_g\ll c$). The above
perturbative expression has been derived by standard first-order
electromagnetic perturbation theory~\cite{Mortensen:2008a} as well
as by a scattering matrix approach combined with the concept of
the Wigner--Smith delay time~\cite{MPedersen:2007a}. So far, the
concept has only been illustrated by means of simulations of ideal
structures comprising lossless dielectric materials infiltrated by
the absorbing bio-liquid of interest. In this paper we illustrate
the effect for lossy materials by means of analytical results for
a Bragg stack composed of a lossy dielectric such as typical
bio-compatible polymer materials envisioned to play an important
role for future low-cost disposable lab-on-a-chip systems.

\section{The Bragg stack example}

The Bragg stack serves as an illustrative example where the
dispersion relation can be obtained by analytical means and where
band diagrams are available also in the case of absorbing
materials. The structure is illustrated in Fig.~\ref{fig1} and is
composed of alternating layers of thickness $b_l$ and $b_s$ with
$\Lambda=b_l+b_s$ being the pitch and
$\epsilon_l=\epsilon_l'+i\epsilon_l''\simeq n_l^2+in_l\alpha_l/k$
and $\epsilon_s=\epsilon_s'+i\epsilon_s''\simeq
n_s^2+in_s\alpha_s/k$ being the corresponding relative dielectric
permittivities of the liquid and the solid. The absorption
coefficients $\alpha_l$ and $\alpha_s$ are defined such that to
lowest order in $\alpha$ the intensity decays exponentially,
$I(z)\propto \big|\exp(i \sqrt{\epsilon}kz )\big|^2=\exp(-\alpha
z)$, in the corresponding spatially homogenous materials.

\subsection{Dispersion relation}

In the following we follow a recent derivation by Kim \emph{et
al.}~\cite{Kim:2006} where the dispersion relation is given by the
root of the following equation
\begin{equation}\label{eq:mastereq}
\cos\left(\kappa\Lambda\right)= F(k),\quad
\end{equation}
where $\kappa$ is the Bloch wave vector and $k(\omega)=\omega/c$
is the free-space wave vector. The dimensionless function on the
right-hand side is given by
\begin{equation}
F(k)=\cos \left(\sqrt{\epsilon_l}k b_l \right) \cos
\left(\sqrt{\epsilon_s}k b_s
\right)-\frac{\epsilon_l+\epsilon_s}{2\sqrt{\epsilon_l}\sqrt{\epsilon_s}}
\sin \left(\sqrt{\epsilon_l}k b_l \right) \sin
\left(\sqrt{\epsilon_s}k b_s \right)
\end{equation}
which is symmetric under the exchange of ($\varepsilon_l,b_l)$ and
($\varepsilon_s,b_s)$.

The group velocity can be found explicitly by noting that
$v_g=\partial\omega/\partial\kappa=c \partial k/\partial\kappa$
from which we get
\begin{equation}\label{eq:vg}
\left(\frac{v_g}{c}\right)^{-1} = \frac{1}{\Lambda}
\frac{\partial}{\partial k}\arccos\left\{ F(k)\right\}=
\frac{1}{\Lambda} \frac{-\frac{\partial}{\partial
k}F(k)}{\sqrt{1-F^2(k)}},
\end{equation}
where it is implicitly assumed that the real part is taken.

\begin{figure}[t!]
\begin{center}
\epsfig{file=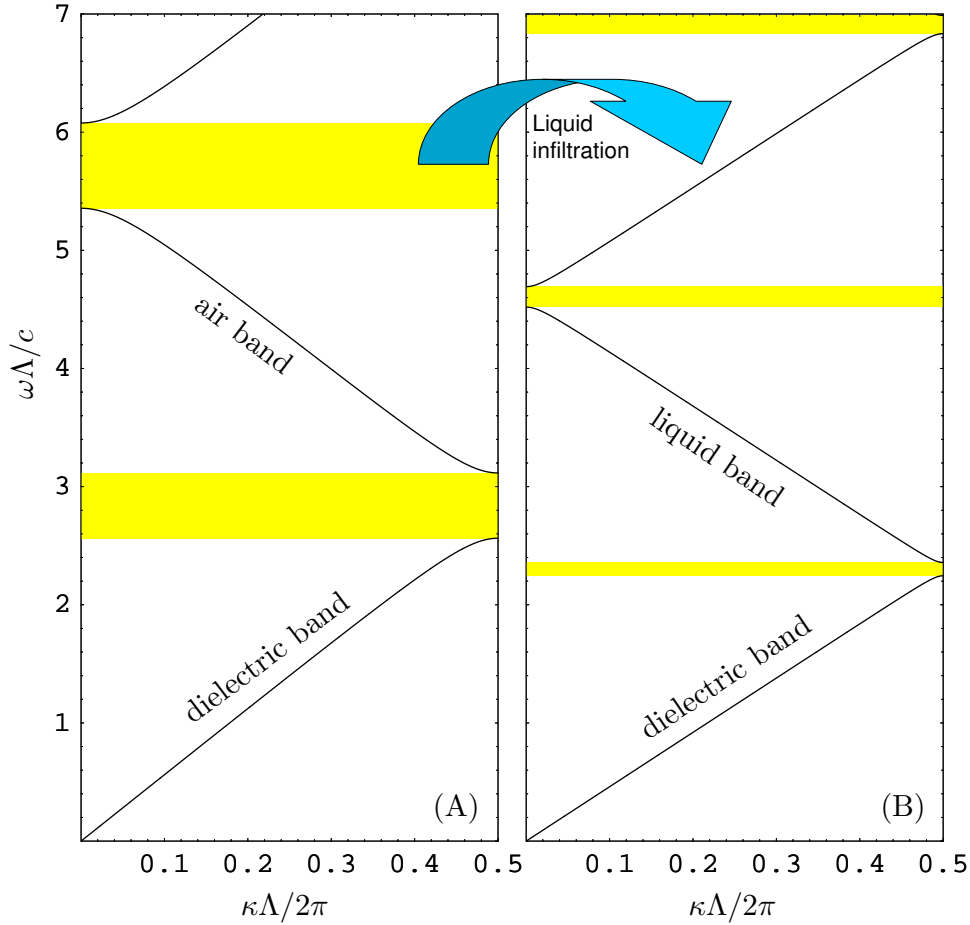, width=0.8\columnwidth,clip}
\end{center}
\caption{(A) Band diagram for a Bragg stack with $n_l=1$
(corresponding to air), $n_s=1.5$ (corresponding to a typical
polymer), $b_l=0.8\Lambda$, and $b_s=0.2\Lambda$. (B) Band diagram
for the structure in (A), but with the air regions infiltrated by
a liquid with $n_l=1.33$.} \label{fig2}
\end{figure}

\begin{figure}[t!]
\begin{center}
\epsfig{file=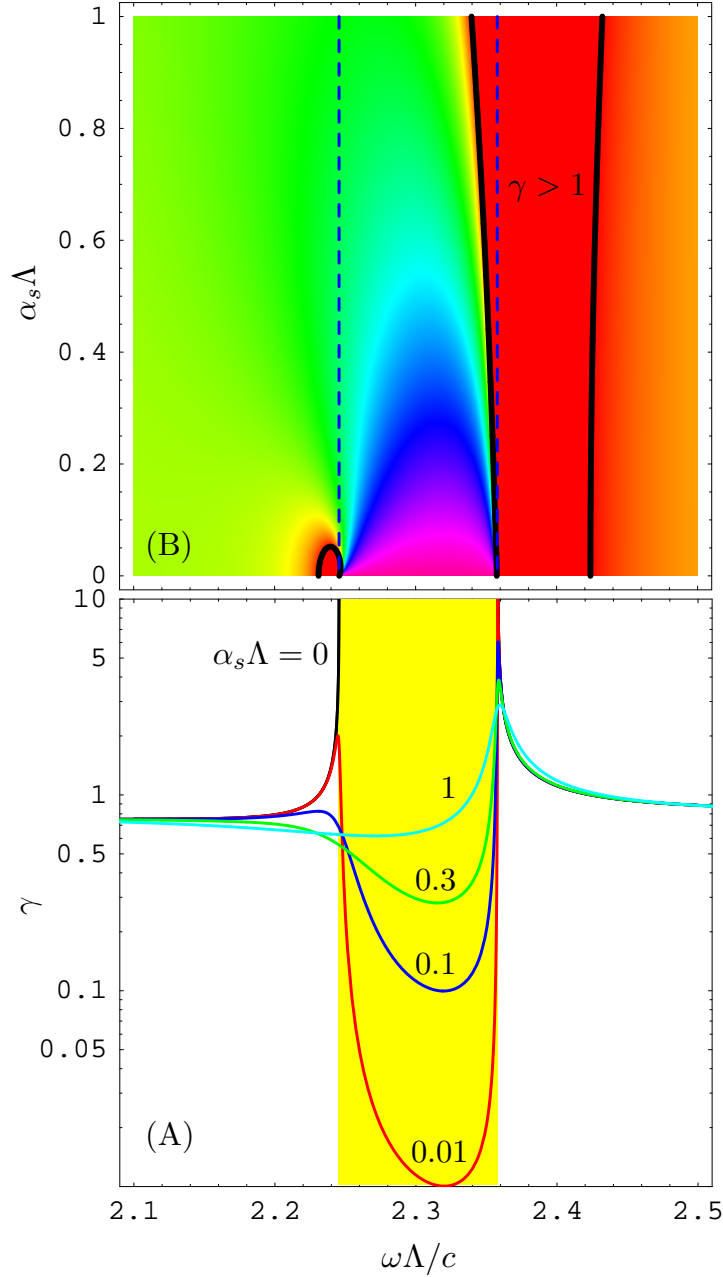, width=0.6\columnwidth,clip}
\end{center}
\caption{Enhancement of liquid absorption for a Bragg stack with
$n_l=1.33$, $n_s=1.5$, $b_l=0.8\Lambda$, and $b_s=0.2\Lambda$. (A)
Beer--Lambert enhancement factor $\gamma$ versus normalized
frequency $\omega\Lambda/c$ around the first band gap (indicated by
yellow shading) for varying values of the absorption
$\alpha_s\Lambda$ in the solid layers. (B) The Beer--Lambert
enhancement factor $\gamma$ in a $\alpha_s\Lambda$ versus
$\omega\Lambda/c$ plot. Regions with $\gamma>1$ are shown in red and
the black solid lines show the contour for $\gamma=1$ while the
dashed blue lines indicate the band-gap edges for the lossless
structure.} \label{fig3}
\end{figure}

In panel (A) of Fig.~\ref{fig2} we illustrate the typical band
structure by means of an example with alternating layers of air
(with $n_l=1$ and $b_l=0.8\Lambda$) and a typical polymer material
(with $n_s=1.5$ and $b_s=0.2\Lambda$). Polymers are often considered
to provide too low an index contrast for the formation of pronounced
photonic band gaps. However, slow-light phenomena in polymer systems
are still feasible as experimentally demonstrated for e.g. photonic
band-edge lasing~\cite{Arango:2007}. When infiltrating the structure
by a typical biological liquid ($n_l=1.33$) the index contrast is
decreased even further and band gaps are expected to narrow. Panel
(B) illustrates this for the same structure as in panel (A), but
with the air regions infiltrated by a liquid with $n_l=1.33$. In
general bands shift down in frequency, but it is noteworthy that
some bands tend to have a more pronounced shift than others. This
has lead to a classification of bands in terms of \emph{liquid} and
\emph{dielectric} bands~\cite{Mortensen:2008a} similarly to the
classification of \emph{dielectric} and \emph{air} bands employed
widely in the photonic crystal community~\cite{Joannopoulos:1995}.
Obviously, the air (liquid) band is most susceptible to the
infiltration by the liquid as seen by the pronounced down-shift in
frequency, while the dielectric band is less perturbed when
introducing the liquid in the air regions. As we shall see, this
different nature of the bands also becomes important for the impact
of loss in the solid layers where the electrical field tends to
localize for the dielectric band.

\subsection{The Beer--Lambert enhancement factor}

In order to calculate the enhancement factor $\gamma$ explicitly
we consider the general case of a complex valued Bloch wave vector
$\kappa=\kappa'+i\kappa''$ with the prime and double-prime
indicating the real and imaginary part, respectively. Taylor
expanding the left-hand side of Eq.~(\ref{eq:mastereq}) to lowest
order in $\kappa''$ and likewise the right-hand side to lowest
order in the liquid absorption parameter $\alpha_l$ we get
\begin{equation}\label{eq:gammaBragg}
\gamma \equiv \frac{2\kappa''}{\alpha_l}= \frac{2}{i\Lambda}\left.
\frac{\frac{\partial}{\partial\alpha_l}F(k)}{\sqrt{1-F^2(k)}}\right|_{\alpha_l=0}.
\end{equation}
This result can be shown to be fully consistent with the general
expression in Eq.~(\ref{eq:gamma}).

In Fig.~\ref{fig3} we illustrate the enhancement for the example
with $n_l=1.33$, $n_s=1.5$, $b_l=0.8\Lambda$, and $b_s=0.2\Lambda$,
corresponding to the band structure shown in panel (B) of
Fig.~\ref{fig2}. We focus on the first band gap centered around
$\omega\Lambda/c\simeq 2.3$, where the band below the band gap is a
dielectric band, while the band just above the band gap is a liquid
band. For the solid layers we include a finite damping quantified by
a varying $\alpha_s$. From the classification of the bands we expect
that the loss will broaden the enhancement in an asymmetric manner
with respect to the center of the band gap, such that the dielectric
band is most susceptible to an increasing $\alpha_s$. Indeed, panel
(A) reveals exactly such a behavior, with a pronounced smearing of
the singularity in $\gamma$ at the lower band edge, while the
singularity at the upper band edge sustains the damping to a higher
degree. This picture is in full agreement with the experimental
observations of enhanced red-absorption in dye-doped inverse-opal
photonic crystals by Nishimura \emph{et al.}~\cite{Nishimura:2003}.
The plot of $\gamma$ in the $\alpha_s$ versus $\omega$ diagram in
panel (B) shows this in more detail. The solid black lines inclose
regions with $\gamma>1$ where the liquid-infiltrated structure is
superior to its classical counterpart without any microstructure.
Interestingly, damping in the solid material is seen not to be a
serious limitation when working with the liquid band. In fact the
damping may even help increase the bandwidth slightly of the
otherwise quite narrow singularity, while still allowing for a
significant enhancement, perhaps even by one order of magnitude.

\section{Conclusion}

An optofluidic slow-light phenomenon has recently been proposed as
a mechanism allowing for enhanced absorption measurements in
lab-on-a-chip systems with a reduced optical path compared to
their macroscopic counterparts. In this work we have explored the
use of lossy dielectric materials such as polymer which is
typically the preferred class of materials for lab-on-a-chip
systems. By means of the Bragg stack example we have explicitly
shown that the enhancement does not critically depend on the
polymer loss and somewhat surprisingly the polymer loss may even
serve as a desired broadening mechanism allowing for a higher
measurement bandwidth.

\section*{Acknowledgments}

This work is financially supported by the \emph{Danish Council for
Strategic Research} through the \emph{Strategic Program for Young
Researchers} (grant no: 2117-05-0037) as well as the \emph{Danish
Research Council for Technology and Production Sciences} (grants
no: 274-07-0080 \& 274-07-0379).

\newpage


\end{document}